\documentstyle[preprint,aps]{revtex}
\begin{document}
\newcommand{\slp}{\slash\hspace{-2mm}p}
\newcommand{\slq}{\slash\hspace{-2mm}q}
\newcommand{\be}{\begin{equation}}
\newcommand{\ee}{\end{equation}}
\draft
\title{ELECTROMAGNETIC PROPERTIES OF OFF-SHELL PARTICLES
   AND GAUGE INVARIANCE}
\author{S.I. Nagorny$^a$, A.E.L. Dieperink$^b$}
\address{$^a$NIKHEF, P.O. Box 41882, 1009 DB Amsterdam, The Netherlands}
\address{$^b$Kernfysisch Versneller Instituut, NL-9747 AA Groningen,
    The Netherlands}
\maketitle
\begin{abstract}
Electromagnetic properties of off-shell particles are discussed on the
sis of a purely electromagnetic reaction: virtual Compton scattering
off a proton. It is shown that the definition of off-shell
electromagnetic form factors is not gauge invariant and they cannot be
investigated in practice. Using only fundamental requirements
of gauge invariance it is demonstrated that off-shell effects are
cancelled in the longitudinal components of the total conserved current
by the ``minimal" contact current, while $off$-$shellness$ appears only in
the transverse (gauge invariant) non-pole part. This provides the
possibility to introduce an $on$-$shell$ $extrapolated$ form factor $F_1^{+}$
in a gauge invariant way for the time-like region $4m_e^2 < q^2 < 4M^2$.
The form factors $F_{1,2}^{-}$ are removed from the total
conserved current and do not affect observables.
\end{abstract}
\section{Introduction}
Electromagnetic (EM) properties of off-shell (virtual)
particles have been the subject of intensive
investigations$^{1-4}$. It was shown that the number of
off-shell form factors (FF) in the general $\gamma^{\star} NN$ vertex
with one (two)- off-shell nucleon(s) increases$^2$ to 4 (6)
compared to two FF for the free nucleon. Dispersion relations$^1$
for off-shell FF were introduced by Bincer. Various one-loop
calculations$^{2-4}$ suggested their model-dependence and possible
influence the observables in the electron-nucleus scattering.
\par
It is well known that any off-mass-shell extension of a matrix element
is not unique: various representations of the field operators give rise
to different off-shell Green functions and off-shell matrix elements,
while all of these lead to the same $S$-matrix. Off-shell properties of
Green functions and in particular their dependence on the representation
of the Lagrangian have been discussed in the framework of field theory for
a long time$^{5}$. Recently the method of effective Lagrangians was applied
to real Compton scattering where the representation dependence of
off-shellness has been pointed out$^{6}$. The possibility to move off-shell
effects from irreducible vertices of the Born current to the regular
part of the amplitude was used$^6$ to obtain a unique definition for
polarizabilities in  Compton scattering on nucleons. Also in
chiral perturbation theory applied to real Compton scattering
off the pion, it was shown$^7$ that off-shell effects depend not only
upon the model of Lagrangian, but also on its representation.
The possibility to shift any explicit off-shell dependence
of the irreducible three-point Green function to the regular part of
the amplitude has also been used to derive$^{8,9}$ the low energy
theorem for the virtual Compton scattering (VCS).
\par
From the general point of view it is clear that the definition of
an $off$-$shell$ FF is directly connected with the possibility to
introduce a $one$-$body$ $off$-$shell$ EM current independently of the
full EM reaction amplitude. However, due to requirements of gauge
invariance one- and many-body currents cannot be considered
separately$^{10}$. Since the one-body off-shell current is not
conserved itself, the definition of off-shell FF will not be gauge
invariant. Therefore, off-shell effects in the $\gamma^{\star} NN$
vertex should be considered in connection with the representation of
the full conserved current.
\par
In the space-like region (and also for $q^2 > 4M^2$), where the on-shell
$\gamma^{\star} NN$ vertex corresponds to a physical process, an
$on$-$shell$ EM FF can be defined in a gauge invariant manner
(they are known, or at least, may be measured), while any off-shell
effects may be considered as corrections and formally transferred to
the non-Born part of the amplitude$^{8,9,11}$.
On other hand, in the so-called ``unphysical"
region ($4m_e^2 < q^2 < 4M^2$)
for any EM process one of the hadrons in the EM vertices is
$off$-$mass$-$shell$ and hence only $off$-$shell$ EM FF exist.
However, their definition cannot be gauge invariant and such a FF
cannot be investigated in practice. That is why in the case of a
virtual time-like photon it is important to understand
whether a representation of the amplitude exists such that the
off-shell effects enter only in the non-pole transverse (gauge
invariant) components of the total conserved current, whereas the
pole-type piece contains only ``on-shell extrapolated" FF.
\par
In general there are various ways (which lead to the
same physical result) to construct a covariant and gauge invariant VCS
amplitude$^{7-9}$. Here we will use a field-theoretical approach$^{10}$
formulated in terms of the n-point Green functions, and explore the
gauge nature of the EM field (associated with the real photon).  The
Ward-Takahashi identities (WTI), which connect the (n+1)-point EM Green
functions (with EM field) and the hadron n-point Green functions
(without EM field), are an important ingredient of the gauge invariant
theory in this case.
\par
In this paper we focus on the off-shell effects associated with $virtual$
photons in the $reducible$ $\gamma^{\star} NN$ vertices which, in the case
of Dirac or scalar particles, do not contain any off-shellness at the
``photon point" due to gauge constraint$^{2,10}$. For our goals we need
not only to introduce an exactly conserved EM current, but it is also
important to define its $irregular$ and $regular$ pieces (pole and contact
current) consistently, from the ``same principle", but without special
dynamical model for off-shellness.
The ``minimal insertion" of the gauge field into n-point Green functions
provides this on the basis of the fundamental requirements only, and it
allows to see how the off-shell effects may be
transferred between one- and many-body reaction mechanisms inside the
total gauge invariant amplitude in a model independent way.
In the ``minimal" scheme the off-shell Born current
(irregular part) and contact current (regular part) will be
defined from the same off-shell $\gamma^{\star} NN$ vertex for which only
the general structure will be assumed.
Of course, such a consideration is restricted
to only the Dirac part of the $\gamma NN$ vertex associated with $real$
$photon$, since its
``abnormal" (Pauli) part cannot be included in the ``minimal scheme"
directly, but in principle it may be considered like a ``vertex
correction" which could be calculated through loop-diagrams.
However, in this case a dynamical model for the higher-order hadron Green
functions (strong vertices) is essential. As for the $\gamma^{\star} NN$
vertices, associated with $virtual$ $photons$, in general, both Dirac and
Pauli parts may formally be included, since we do not suppose ``minimal"
procedure in this case and we can use the most general
phenomenological expression satisfying fundamental requirements. However,
for simplicity we will restrict ourselves to only the Dirac FF in the
$\gamma^{\star} NN$ vertex also, and mention additional effects associated
with the Pauli FF if necessary.
\par
We note that the present study differs from  previous studies of the
VCS$^{8-9,11}$ (for completeness see also the refs. in [9])
were based upon the introduction of an ``on-shell Born current",
transferring all possible off-shell effects to the unknown
regular part; the latter  was then parametrised in terms of the invariant
functions - ``generalized polarizabilities" - on the basis of the
fundamental requirements of covariance, CPT and gauge invariance.
\par
The paper is arranged in following way. In Section II a purely EM process
$\gamma + p \Longleftrightarrow \gamma^{\star} + p$ (space- or time-like
VCS off a proton) is considered. First, we give the definitions of the
various hadrons and EM Green functions (without and with EM field), as
well as corresponding WTI to which the last ones should satisfy. Then,
using minimal substitution of the gauge field into the corresponding Green
functions, the structure of the $off$-$shell$ Born current,
the contact current and the total conserved VSC current is obtained
on the basis of the same principle, in terms of
the $reducible$ vertices and $free$ Feynman propagators.
We show that $off$-$shell$ form factors cannot be defined in
a gauge invariant way, even if $\gamma^{\star} NN$ vertex satisfy
corresponding WTI,
and, as a result, they cannot be investigated in
reality.
\par
In Section III we deal with the "off-shellness" in the $regular$
and $irregular$ pieces of the conserved current. In a model
independent way we show that off-shell effects, which appear
only for the $virtual$ photons in the $reducible$ $\gamma^{\star} NN$
vertices, may be not only moved from the pole-part of the amplitude to
its regular part, but also $cancelled$ in the charge operators
of the total conserved current, if the pole and contact amplitudes are
considered consistently. Additionally, we will see that form
factor $F^{-}_{1,2}$, connected with negative energy states of the virtual
proton, will be cancelled in the total conserved current and will not
influence the observables.
\par
Section IV is devoted to an alternative construction of the total
conserved current in terms of the $irreducible$ vertices and $full$
renormalized propagators, again on the basis of the minimal
substitution procedure. Another representation of the ``minimal"
conserved current for VCS is obtained there.
The relation between two different representations of the conserved
current, based on the $reducible$ and $irreducible$ $\gamma^{\star} NN$
vertices, is established. In the framework of the ``minimal" scheme,
using general properties of the mass-operator, it was shown that the
``corrections" to the Born and contact currents, stipulated by the
self-energy part, have the same absolute value but opposite signs. So,
they cancel one another in the total conserved current in a such a model
independent way that the amplitudes defined in terms of the $reducible$
vertices with $free$ Feynman propagators and $irreducible$ ones
together with $full$ renormalized propagators become the same. At last,
for the $real$ photons the off-shellness in the $irreducible$ vertices
may be isolated incide the ``correction terms" only and, due to their
cancellation in the total current, may be not considered at all. This
simply corresponds to the transition to the representation of the
conserved current in terms of the $reducible$ vertices which are free
from the off-shellness due to the gauge constraint.
\par
Section V is devoted to an application of the gauge invariant
definition of the ``on-shell extended" form factors in the time-like
$unphysical$ region, which is only possible due to special
representation of the total amplitude, and is based on the cancellation
of the off-shellnes in the conserved current. The sensitivity of the
$e^{+}e^{-}$- asymmetry for the di-lepton production off the proton
in a ``special" kinematics to such a ``gauge invariant" form factors is
considered. In Section VI we discuss the results and give the
conclusions. Some details and intermediate transformations are given
in the Appendices.
\section{\bf Structure of the VCS amplitude}
In this section the structure of the gauge invariant nucleon current
for VCS off a proton with accounting for the off-shell effects in the EM
vertices will be considered. For a Dirac proton (in relation to the $real$
$photon$) or scalar particles a consistent definition of the regular and
irregular pieces of the total conserved current, based on the gauge nature
of EM field, is possible without a special model for the off-shell behaviour
of the FF. To derive this, let us introduce first strong and EM Green's
functions, as well as the corresponding WTI to which the lastes should satisfy.
\subsection{Preliminaries}
The two-point Green function of the proton, as well as the
electromagnetic three- and four-point Green functions are defined as
$$ S(x,y) = -i <0|T\{\psi(x) \bar{\psi}(y)\}|0>, $$ 
$$G_{\mu}(x,y,z) = <0|T\{\psi(x) \bar{\psi}(y) j_{\mu}(z)\}|0>,$$
$$ G_{\mu \nu}(x,y,z,r) = <0|T\{\psi(x) \bar{\psi}(y) j_{\mu}(z)
j_{\nu}(r)\}|0>, $$ 
where $j_{\mu}(r) = -\nabla^2 A_{\mu}(r)$ is the EM current operator,
 $\psi(x)$ denotes an interpolating field operator of the proton,
and $A_{\mu}$ is EM field satisfying Lorentz condition $\nabla_{\mu} A_{\mu}
(r) = 0$; $T$ denotes the time-ordered product.
\par
Using translation invariance the corresponding momentum-space
Green functions are
$$ (2 \pi)^4 \delta^4(p-p') S(p) = \int d^4 x d^4 y e^{i(px -p'y)}
S(x,y), $$
$$(2 \pi)^4 \delta^4(p-p'-q') G_{\mu}(p,p') = \int d^4 x d^4 y d^4 r
e^{i(px - p'y - q'r)} G_{\mu}(x,y,r), $$
$$ (2 \pi)^4 \delta^4(p+q-p'-q') G_{\mu}(P,q,q') = \int d^4 x d^4 y d^4 r
e^{i(px + qz - p'y - q'r)} G_{\mu}(x,y,z,r), $$
where $P=p+p'$, and $q$, $q'$ are the photon momenta assosiated with
currents $j_{\mu}$ and $j_{\nu}$.
\par
The truncated three- and four-point electromagnetic Green functions, or
$\gamma NN$ and $\gamma NN \gamma$ vertices
are obtained by ``amputation" of the external proton lines:
$$\left \{ \begin{array}{c} \Gamma_{\mu}(p,p') \\ \Gamma_{\mu \nu}
(P,q,q') \end{array} \right \} \ \ \ = \ \ \ S^{-1}(p) \left \{
\begin{array}{c} G_{\mu}(p,p') \\ G_{\mu \nu}(P,q,q') \end{array}
\right \} S^{-1}(p')$$
To be more concrete from now on we will distinguish the half-off-shell
$reducible$ $\gamma NN$ vertex ($\Gamma_{\mu}(p,p')$) and
the $irreducible$ one ($\Gamma_{\mu}^{\rm ir}(p,p')$)
\be {S(p) \Gamma_{\mu}^{\rm ir}(p,p') S(p') = S_{0}(p) \Gamma_{\mu}(p,p')
S_{0}(p')}, \label{eqno(1)} \ee
where $S(p)$ and $S_{0}(p)$ is the full renormalized and free Feynman
propagators.
\par
In the case of $half$-$off$-$shell$ $\gamma NN$ ir-reducible or reducible
vertices (with real or virtual photons) the corresponding WTI have the
same form  (see Appendix A):
\be {q_{\mu} \Gamma_{\mu}^{\rm ir}(p,p')=e\{S^{-1}(p)-S^{-1}(p')\}},
\label{wtia} \ee
\be {q_{\mu} \Gamma_{\mu}(p,p')=e\{S_{0}^{-1}(p)-S_{0}^{-1}(p')\}},
\label{wtib} \ee
\subsection{Conserved current in terms of reducible vertices}
A conserved current for a purely EM reaction
$\gamma + p \Longleftrightarrow \gamma^{\star} + p$
on a Dirac proton can be obtained using
minimal substitution of the EM field into the three-point Green
function$^{10}$, corresponding to the $\gamma^{\star} NN$ vertex (with
virtual photon). In general  this procedure may be carried out  in
two different ways, namely on either the right-hand or the
left-hand side of eq.(\ref{eqno(1)}).
In the first case the theory will be formulated in terms of the
$reducible$ vertices and free Feynman propagators, while the second case
will lead to the formulation in terms of
$irreducible$ vertices and full renormalized propagators. Even from the
beginning it is evident that the same results for physical observables
should be obtained, since a gauge (massless) field is ``inserted"
into the different sides of identity.
In this section we start with the right-hand side of eq.(\ref{eqno(1)})
and therefore we concentrate on the off-shellness in the $reducible$
vertices.
(The latter occurs only for virtual photons and does not exist
 at the ``photon point".)
\par
The minimal insertion of the second (real) photon into the external
lines of the right-hand side of eq.(\ref{eqno(1)}) generates$^{10}$ the one-body
$off$-$shell$ Born current (see Appendix B)
\be {J^{B-off}_{\mu \nu} = e \{ \Gamma_{\mu}(p,p'+q') S_{0}(p'+q')
\gamma_{\nu} + \gamma_{\nu} S_{0}(p-q') \Gamma_{\mu}(p-q',p') \} }.
\label{jbornoff0} \ee
Here $p'$($p$) is initial (final) momentum of the nucleon, $q$($q'$) is
the virtual (real) photon momentum: $p+q=p'+q'$; $e$ is a charge, and
$\gamma_{\nu}$ is a 4$\times$4 Dirac matrix.
\par
The minimal photon insertion inside the $\gamma^{\star} NN$ vertex gives
rise to the contact (``many-body") current$^{10}$,
containing only one-body irreducible graphs (diagrams which cannot be
disconnected by cutting only any single-particle internal lines). Note,
that the requirement of gauge invariance itself does not fix the contact
current completely, since it still depends
on the ``trajectory"$^{10}$. The additional assumption of the $minimal$
insertion of the EM field, corresponding to the ``minimal (linear)
trajectory", makes its definition unique$^{10}$
\be {J^{C}_{\mu \nu} = - e \int^1_0 {d \lambda \over \lambda}
{d \over d q'_{\nu}} \{ \Gamma_{\mu}(p, p'+\lambda q') - \Gamma_{\mu}
(p-\lambda q',p') \} }. \label{jc0} \ee
It is easy to check that the derivative of $J^C_{\mu \nu}$  cancels
corresponding derivative of $J^{B-off}_{\mu \nu}$ hence the
total nucleon current
\be {J^{N}_{\mu \nu} = J^{B-off}_{\mu \nu} +
J^{C}_{\mu \nu}} \label{jn0} \ee
is conserved for any off-shell $\gamma^{\star} NN$ vertex, independent of
its explicit form (initial and final spinors are implied here)
$$ {q_{\mu}J^{N}_{\mu \nu} = J^{N}_{\mu \nu}q'_{\nu} = q_{\mu}
(J^{B-off}_{\mu \nu} + J^{C}_{\mu \nu})=
(J^{B-off}_{\mu \nu} + J^{C}_{\mu \nu})q'_{\nu} = 0}. $$
We note that $J^{B-off}_{\mu \nu}$ and $J^{C}_{\mu \nu}$ are not gauge
invariant separately
$${J^{B-off}_ {\mu \nu} q'_{\nu} =
 - J^{C}_ {\mu \nu} q'_{\nu} =
e  \{ \Gamma_{\mu}(p,p'+q') -
\Gamma_{\mu}(p-q',p') \} \neq 0}, $$
although $q_{\mu} J^{B-off}_{\mu \nu} = q_{\mu}
J^{C}_{\mu \nu} = 0 $.
Only the $on$-$shell$ Born current, $J^{B-on}_{\mu \nu},$ is conserved$^8$,
since the corresponding vertex depends only upon the photon momentum:
$\Gamma_{\mu}^{on}(p,p') \equiv \Gamma_{\mu} (p-p')$.
\par
The crossing properties of the total current eq.(\ref{jn0}) are considered
in Appendix C.
\subsection{Gauge dependence of the definition of off-shell EM FF}
Since the one-body off-shell current is not conserved (see above), the
definition of the off-shell FF is not gauge invariant in spite of the
fact that the half-off-shell $\gamma^{\star} NN$ vertex satisfies the WTI
for the three-point Green function. Indeed, if one of the nucleons is
off-mass-shell, the vertex $\gamma^{\star} NN$ is only a part of the
one-body reducible graph (nucleon-pole diagram). The corresponding one-body
off-shell current cannot satisfy the next order WTI (for the Green function
of the considered process, for example, the four-point one) itself, without
a contribution of the ``many-body" (contact) current, i.e.
the regular piece of the amplitude. Therefore, the off-shell
$\gamma^{\star} NN$ vertex cannot be considered independent
of the full conserved current and the definition of the off-shell FF
cannot be gauge invariant.
\par
In practice the contribution of the off-shell Born current
$J^{B-off}_{\mu \nu}$, which contains the off-shell EM FF, to the
amplitude and any observables is proportional to the contraction:
\be {T^{g}(P,q,q') \ \ \sim \ \ j^{lept}_{\alpha}(q) \  D_{\alpha \mu}^{(g)}(q^2) \
J_{\mu \nu}^{B-off}(P,q,q') \epsilon_{\nu}(q')}, \label{gauge1} \ee
where $D_{\alpha \mu}^{(g)}(q^2)$ is the propagator of the virtual photon and
index $g$ denotes its gauge;
$\epsilon_{\nu}(q')$ is the polarization vector of the real photon, and
$j_{\alpha}^{lept}(q)$ is a leptonic current which corresponds
to the $e^{-} \gamma^{\star} e^{-}$ ($\gamma^{\star} e^{+}e^{-}$) vertex for
space-like (time-like) photon, and which is conserved automatically
(one-photon approximation)
\be {q_{\alpha} \ j_{\alpha}^{lept}(q) \ = \ 0 },
\label{gauge2} \ee
since the initial/final leptons are on-mass shell.
\par
Let us consider the sensitivity of the amplitude $T^{g}(P,q,q')$, eq.(\ref{gauge1}),
to the choice of the gauge for the virtual photon propagator.
In general, any known gauge can be used for $D^{(g)}_{\alpha \mu}(q^2)$, but here
we will consider only few the most common of them, namely the
covariant Feynman ($\eta = 1$) and Landau ($\eta = 0$) gauge
\be {D_{\alpha \mu}^{(g)}(q^2) = {1 \over q^2 + i 0} \{ g_{\alpha \mu} +
(1- \eta) {q_{\alpha} q_{\mu} \over q^2} \} }, \label{gauge3} \ee
and ``axial gauge"
\be {D_{\alpha \mu}^{(g)}(q^2) = {1 \over q^2 + i 0} \{ g_{\alpha \mu} +
{q_{\alpha} q_{\mu} \over (n q)^2} n^2 - {n_{\alpha} q_{\mu} +
q_{\alpha} n_{\mu} \over (n q)} \} }, \label{gauge4} \ee
where $n^2 = \pm 1, 0$ corresponds to ``axial" ($A_3 = 0$), Weyl
($A_{0} = 0$) and ``light-cone" ($A_0 - A_3 = 0$) gauges.
Substituting eq.(\ref{gauge3}) into eq.(\ref{gauge1}) and taking into
account eq.(\ref{gauge2}), we see that the contribution of the
$J_{\mu \nu}^{B-off}$ to the amplitude $T^{g}(P,q,q')$
has a $fixed$ $value$ which is the same for both covariant gauges
(Feynman and Landau), since the leptonic current is conserved
$$ {T^{\eta = 0}(P,q,q') = T^{\eta =1}}(P,q,q'). $$
However, in general the conservation of only the leptonic current in
eq.({\ref{gauge1}) is not enough for a gauge invariant definition of the
off-shell EM FF. Indeed, substituting eq.(\ref{gauge4}) into
eq.(\ref{gauge1}) and taking into account that $n_{\alpha}
j_{\alpha}^{lept} \neq 0$, we see that for any ``axial" gauge the
amplitude $T^g(P,q,q')$ differs from
the ``covariant" ones. Moreover, various
``axial" gauges, which are $equivalent$ in principle, will also lead to
different results
$$ {T^{n^2 = 1}(P,q,q') \ \neq \ T^{n^2 = -1}(P,q,q') \ \neq \
T^{n^2 = 0}(P,q,q')}. $$
Therefore, the definition of the off-shell EM FF in the one-body current,
even through $\gamma^{\star} NN$ vertex which satisfy the WTI for the
three-point EM Green function, is gauge
dependent and such a form factors cannot be investigated in practice.
Only exactly conserved hadron currents, like $J_{\mu \nu}^{N}$
from eq.(\ref{jn0}), may lead to $physical$ contributions to $observables$.
\section{\bf Gauge Invariant Decomposition of The Total Nucleon
Current}
The reducible half-off-shell $\gamma^{\star} NN$- vertex,
satisfying the WTI (\ref{wtib}), contains two terms$^1$
$\Gamma^{\pm}_{\mu}(q^2,p^2 \neq M^2)$
 corresponding to the positive and negative
energy states of the virtual nucleon
\be {\Gamma_{\mu} = \Gamma_{\mu}^{+}(q^2,p^2) \ {\slp + M \over 2M} +
\Gamma_{\mu}^{-}(q^2,p^2) \ {- \slp + M \over 2M}},
\label{redvertex} \ee
which can be expressed$^{1-3}$ through the
$off$-$shell$ EM FF $F_{1,2}^{\pm}(q^2,p^2)$:
\be
{\Gamma_{\mu}^{\pm}(q^2,p^2) = e \{F_{1}^{\pm}(q^2,p^2) \gamma_{\mu} + {1
- F_{1}^{\pm}(q^2,p^2) \over q^2} \slq q_{\mu} - {\sigma_{\mu \nu} q_{\nu}
\over 2 M} F_2^{\pm}(q^2,p^2) \}},
\label{pmredvertex}  \ee
where $\sigma_{\mu \nu} = (\gamma_{\mu} \gamma_{\nu} - \gamma_{\nu}
\gamma_{\mu})/2$, and $\slq = \gamma_{\mu} q_{\mu}$.
\par
For an on-mass-shell particle only $\Gamma^{+}_{\mu}
(q^2,p^2=M^2)$ exists, and $F^{+}_{1,2} (q^2,p^2)$ may be considered as
the $physical$ FF of the nucleon, contrary to $F_{1,2}^{-}(q^2,p^2)$.
\par
For simplicity we will not consider the Pauli FF $F^{\pm}_2$ in
eq.(\ref{pmredvertex}) explicitly. However, we stress that the procedure,
similar to that one described below, may be directly apply to the full
vertex in eq.(\ref{pmredvertex}).
\par
At the photon point gauge invariance demands$^{2,10}$
$F_{1}^{\pm}(q^2 = 0, p^2) \ \ \equiv \ \ 1$.
Therefore, assuming a regular behaviour of the $physical$ FF
as a function of $q^2$ and $p^2$
(which is evident  below the pion threshold), we get a
model independent representation
\be
{F_{1}^{+}(q^2, p^2) = F_1^{+} (q^2,M^2) +
{q^2 \over M^2} {p^2-M^2 \over M^2 } A^{+}(q^2,p^2) }.
\label{ffpl} \ee
Here the analytical function $A^{+}(q^2,p^2)$ contains all off-shell effects
\be {A^{+}(q^2,p^2) = \sum_{i,k=1}^{\infty}
a_{i k}^{+} \  ({p^2-M^2 \over M^2})^{i-1} ({q^2 \over M^2})^{k-1}},
\label{apl}
\ee
and $a_{i k}^{+}$ are constants, corresponding to the $i,k$- order
derivatives at $p^2=M^2$, $q^2=0$.
\par
To derive a relation between $off$- and $on$-$shell$ Born currents, we
substitute eqs.(\ref{redvertex},\ref{pmredvertex}) into
eq.(\ref{jbornoff0}).
Using eq.(\ref{ffpl}) and taking into account$^{6,9}$ that the numerator
$p^2-M^2$ cancels the denominators of the propagators in
eq.(\ref{jbornoff0}),
as well as the identities: $(\slp +M) S_{0}(p) = 1 + 2M S_{0}(p)$ and
$(-\slp+M) S_{0}(p) = - 1$, we express $J_{\mu \nu}^{B-off}$ from
eq.(\ref{jbornoff0}) in terms of the gauge invariant
$on$-$shell$ Born current $J_{\mu \nu}^{B-on}$ (containing only $pole$-$type$
singularities), and a regular term
(containing only $one$-$body$-$irreducible$ contributions); the latter
consists of a gauge invariant part, $J^{GI}_{\mu \nu}$, and a non-gauge
invariant one, $J^{non-GI}_{\mu \nu}$
\be
{J_{\mu \nu}^{B-off} = J_{\mu \nu}^{B-on} + J_{\mu \nu}^{GI} +
J_{\mu \nu}^{non-GI} }. \label{jbornoff} \ee
The explicit form of the various currents in eq.(\ref{jbornoff}) is
\be
{J_{\mu \nu}^{B-on} = e^2 \{\Gamma_{\mu}^{on} S_{0}(p'+q') \gamma_{\nu} +
\gamma_{\nu} S_{0}(p-q') \Gamma_{\mu}^{on}\}}. \label{jbornon}  \ee
\be {J_{\mu \nu}^{GI} = {e^2 q^2\over M^4} \{A^{+}(q^2,W^2)
\tilde{\gamma}_{\mu} \slq' \gamma_{\nu} - \gamma_{\nu} \slq'
\tilde{\gamma}_{\mu} A^{+}(q^2,W'^2)\}}.\label{jgi} \ee
\be J_{\mu \nu}^{non-GI} = \frac{2 e^2 q^2}{ M^4} \{p'_{\nu} A^{+}(q^2,W^2)
+ p_{\nu} A^{+}(q^2,W'^2)\} \tilde{\gamma}_{\mu}
  + \frac{e^2}{ 2M} \{\Delta(W^2) \tilde{\gamma}_{\mu} \gamma_{\nu} +
\gamma_{\nu} \tilde{\gamma}_{\mu} \Delta(W'^2)\}. \label{jnongi} \ee
Here $\Delta(W^2)=F_1^{+}(q^2,W^2)-F_1^{-}(q^2,W^2)$,
$W^2 = (p'+q')^2$, $W'^2 = (p-q')^2$, and
\be \Gamma_{\mu}^{on} = \Gamma_{\mu}^{+}(q^2, M^2)
\ \ ; \ \ \ \ \ \ \ \ \
\tilde{\gamma}_{\mu} = \gamma_{\mu} - {\slq q_{\mu} \over q^2} \nonumber \ee
\par
From eq.(\ref{jbornoff}) we see that $off$-$shell$ $effects$ in the $\gamma^{\star}
NN$- vertex may be completely moved (see also [6-9]) to the regular piece of the
amplitude (\ref{jgi},\ref{jnongi}), containing only irreducible graphs. This statement
is general and does not depend upon the type of the EM process (it may be
applied directly to the photo/electro- disintegration of a bound system).
\par
Using eqs.(\ref{jbornon}-\ref{jnongi}), it is easy to see that every current
in eq.(\ref{jbornoff}) has
a non-singular limit at $q^2 \rightarrow 0$, and in the photon point
(real Compton scattering off a Dirac or scalar particle) off-shell
effects vanish (as a consequence of gauge invariance)
$$J^{B-off}_{\mu \nu} (q^2=0) \equiv J^{B-on}_{\mu \nu}(q^2=0)$$
\be
{J^{GI}_{\mu \nu}(q^2=0) \ = \ J^{non-GI}_{\mu \nu}(q^2=0) \ = \ 0}.
\label{jgi0} \ee
To calculate the contribution of the ``minimal" contact current  we
substitute eq.(\ref{redvertex}) into eq.(\ref{jc0}). Then, taking into account requirements of
covariance, the integration in eq.(\ref{jc0}) can be performed analytically, independent
of the explicit form of the vertex function
$$
{-{1 \over e} J_{\mu \nu}^{C} = [\Gamma_{\mu}^{+}(W^2)-
\Gamma_{\mu}^{-}(W^2)] \gamma_{\nu}+
\gamma_{\nu} [\Gamma_{\mu}^{+}(W'^2) - \Gamma_{\mu}^{-}(W'^2)]  }$$
\be {+ \ {p'_{\nu} \over p'q} [\Gamma_{\mu}^{+}(W^2)-\Gamma_{\mu}^{+}(M^2)] -
{p_{\nu} \over pq} [\Gamma_{\mu}^{+}(W'^2)-\Gamma_{\mu}^{+}(M^2)]}.
\label{jc1} \ee
Substituting eqs.(\ref{pmredvertex},\ref{ffpl})  into eq.(\ref{jc1}), we
find that the contact current (\ref{jc1}) exactly coincides
with the non-gauge invariant piece of
the regular term (\ref{jnongi}), but with opposite sign
\be {J_{\mu \nu}^{C} = - J_{\mu \nu}^{non-GI}  }. \label{jc2} \ee
Summing eqs.(\ref{jc2}) and (\ref{jbornoff}),
 we see that the ``minimal" contact current (\ref{jc2})
kills all non-gauge invariant terms in $J_{\mu \nu}^{B-off}$, and
as a result, off-shell effects are completely cancelled in the
``longitudinal" part (in the charge operators) of the total conserved
nucleon current $J_{\mu \nu}^{N}$
\be {J_{\mu \nu}^{N} = J_{\mu \nu}^{B-off} + J_{\mu \nu}^{C} = J_{\mu \nu}
^{B-on} + J_{\mu \nu}^{GI} },\label{jn} \ee
where $q_{\mu} J_{\mu \nu}^{B-on} = J_{\mu \nu}^{B-on} q'_{\nu} = q_{\mu}
J_{\mu \nu}^{GI} = J_{\mu \nu}^{GI} q'_{\nu} =0$, and only the $new$
``transverse" contact current $J_{\mu \nu}^{GI}\sim \tilde{\gamma}_{\mu}
\sigma_{\nu \alpha} q'_{\alpha}$, caused by the Dirac magnetic moment,
depends on off-shell effects. In case of VCS off a scalar particle the
cancellation of off-shell effects is complete.
As it may be seen from eqs.(\ref{jgi},\ref{jnongi}), the expansion
(\ref{apl}) was not used to derive eq.(\ref{jn}).
\par
Therefore, we obtained a novel decomposition (\ref{jn}) of the total conserved
nucleon current into two gauge invariant pieces $J_{\mu \nu}^{B-on}$
and $J_{\mu \nu}^{GI}$
which shows:  1) the introduction of $off$-$shell$ FF in the ``reducible"
$\gamma^{\star} NN$ vertex is not gauge invariant and depends on the
representation of the total conserved current; 2)
off-shell effects may be distributed over the $\gamma^{\star}NN$- vertex
and the contact current in different ways; 3) off-shell effects from one-body
currents are completely cancelled in the longitudinal components of the
total conserved current by the ``minimal" contact current;
4) $F_1^{-}$ is removed from the total conserved
nucleon current and does not influence the observables.
Therefore, a representation of the total conserved current such that the
off-shell effects enter only in the $non$-$pole$ gauge invariant piece and
only through Dirac magnetic operators exists
for both space- and time-like regions. This enables one to introduce an
``on-shell extrapolated" FF for the ``unphysical" region in a gauge
invariant way.
\par
Note, that inclusion of the $F_2^{\pm}$ form factors in the same manner
does not change the structure and the physical meaning of eq.(\ref{jn}).
This leads just to only an additional term in $J^{GI}_ {\mu \nu}$,
containing the physical FF $F_2^{+}$ only, proportional to:
$${ \left ( {d \over d p^2} F_2^{+}(q^2,p^2) \right )_{p^2=M^2}
\{\sigma_{\mu \alpha}
\sigma_{\beta \nu} + \sigma_{\beta \nu} \sigma_{\mu \alpha} \} q_{\alpha}
q'_{\beta} }. $$
There are also  two additional terms in $J_{\mu \nu}^{non-GI}$,
including both $F_2^{\pm}$ form factors; we do not present these terms here,
since they are cancelled by the
corresponding terms from the ``minimal" contact current, in the same
manner as was shown above for $F^{\pm}_1$. Therefore, we can conclude,
that the form factor $F_2^{-}$ is also removed from the total conserved
current, similary to $F_1^{-}$, and does not influence the observables.
\par
We stress that the real Compton scattering amplitude
off a Dirac (or scalar) particle cannot contain any off-shell
effects at all, as a consequence of gauge invariance.
\par
Finally, all results and conclusions, obtained above, are
valid for both space/time-like regions, and the definition of the
$on$-$shell$ $Born$ current is $unique$ under the condition that
corresponding $\gamma^{\star} NN$ vertices satisfy the WTI on the
$operator$ $level$ (this is always true for a $Dirac$ proton).
\section{\bf Conserved Current in Terms of Irreducible Vertices}
Here we will briefly repeat the same procedure of the ``minimal
substitution" of the EM field in the $left$-$hand$ side of (\ref{eqno(1)}),
which contains full propagators and an irreducible vertex. In this
case minimal insertion of the second (real) photon into external lines
(full renormalized nucleon propagators) generates$^{10}$, similar to
eq.(\ref{jbornoff0}), an alternative
expression for the one-body $off$-$shell$ Born current
\be
 \tilde{J}^{B-off}_{\mu \nu} =  \Gamma_{\mu}^{\rm ir}(p,p'+q')
S(p'+q') \Gamma_{\nu}^{ir}(p'+q',p') +  \Gamma_{\nu}^{ir}(p,p-q')
S(p-q') \Gamma_{\mu}^{\rm ir}(p-q',p') .
\label{jtbornoff} \ee
Note, eq.(\ref{jtbornoff}) is formally the same as the one used in ref
[9] for the contribution of the ``class A" diagrams.
However, in this paper the $irreducible$ $\gamma NN$ vertex (with a
real photon) is now the ``minimal" three-point EM Green's function
connected with inverse propagator$^{10}$ (see also [12])
\be {\left \{ \begin{array}{c} \Gamma_{\nu}^{ir}(p'+q',p') \\
\Gamma_{\nu}^{\rm ir}(p,p-q') \end{array} \right \} \ \ \ = \ \ \ \pm e
\int_{0}^{1} {d \lambda \over \lambda}{d \over d q'_{\nu}} \ \ \left \{
\begin{array}{c} S^{-1}(p'+\lambda q') \\ S^{-1}(p-\lambda q')
\end{array} \right \}}, \label{gammair4} \ee
and, as a result, satisfies the WTI (\ref{wtia}) automatically.
Eq.(\ref{gammair4}) is a consequence$^{10}$ of the ``minimal coupling" of
the second (real) photon to the full renormalized propagator, and
follows from the eqs.(21)-(22) in the first ref. [10].
The $irreducible$ $\gamma^{\star} NN$ vertices (with $virtual$ photons),
$\Gamma_{\mu}^{\rm ir}(p'+q',p)$ and $\Gamma_{\mu}^{\rm ir}(p-q',p'),$
satisfy the  WTI (\ref{wtia}) also$^{13}.$
However, for these we do
not need a  "minimal" procedure,  and their Lorentz-spin structure
may be described by the same equations like eqs.(\ref{redvertex},
\ref{pmredvertex}).
\par
The minimal photon insertion inside the $irreducible$
$\gamma^{\star} NN$ vertex gives rise the new contact current which
has the same structure as eq.(\ref{jc0}); however it  contains
$irreducible$ vertices instead of $reducible$ ones
\be
{\tilde{J}^{C}_{\mu \nu} = - e \int^1_0 {d \lambda \over \lambda}
{d \over d q'_{\nu}} \{ \Gamma_{\mu}^{\rm ir}(p, p'+\lambda q') -
\Gamma_{\mu}^{\rm ir}(p-\lambda q',p') \} }. \label{jtc} \ee
Using the WTI (\ref{wtia}) and eqs.(\ref{gammair4}), as well as taking into
account that $\bar{u}(p) S^{-1}(p) = \bar{u}(p) S^{-1}_{0}(p)=
S^{-1}(p') u(p') = S^{-1}_{0}(p') u(p') =0$, it is easy to get convinced
that the derivatives of $\tilde{J}_{\mu \nu}^{C}$ cancel corresponding
derivatives of $\tilde{J}_{\mu \nu}^{B-off}$, independent of the
explicit form of the vertices, and the $new$ total nucleon
current
\be {\tilde{J}^{N}_{\mu \nu} \ \ \ = \ \ \ \tilde{J}_{\mu \nu}^{B-off}
\ \ \ + \ \ \ \tilde{J}^{C}_{\mu \nu}} \ ,\label{jtn} \ee
defined in the terms of the $irreducible$ vertices and $full$
renormalized propagators, is exactly conserved
$$ {q_{\mu} \tilde{J}^{N}_{\mu \nu} \ \ = \ \ \tilde{J}_{\mu \nu}^{N}
q'_{\nu} \ \ = \ \ 0}. $$
To establish a relation between two gauge invariant nucleon
currents, $J_{\mu \nu}^{N}$ from (\ref{jn0})
 and $\tilde{J}_{\mu \nu}^{N}$ from (\ref{jtn}),
 we have to take
into account the connection between ``full" renormalized and ``free"
Feynman propagators
\be {S^{-1}(p) = \slp + M - \Sigma(p) \ \ \ ; \ \ \ \ S^{-1}_{0}=
\slp + M  \ \ \ ; \ \ \ \ S^{-1}_{0}(p) - S^{-1}(p) \ =
\ \Sigma(p)},\label{smin1} \ee
where the mass-operator $\Sigma(p)$ satisfies the following conditions:
\be {\Sigma(p) |_{p^2=M^2} = 0 \ \ \ ; \ \ \ \ {d \over d \slp}
\Sigma(p) |_{p^2=M^2} =0}. \label{mass} \ee
Substitution of eq.(\ref{smin1}) into eq.(\ref{gammair4})
 gives the connection between
the ``minimal" $\gamma NN$ $irreducible$ vertex and the $reducible$ one
\be
{\Gamma_{\nu}^{\rm ir}(p'+q',p')  =  e \gamma_{\nu} - e \int^{1}_{0}
{d \lambda \over \lambda}
{d \over d q'_{\nu}} \Sigma(p'+\lambda q')},
\label{gammair3} \ee
Using eq.(\ref{gammair3}) and taking into account that $\Gamma_{\mu}^{\rm ir}
(p,p'+q') S(p'+q') = \Gamma_{\mu}(p,p'+q') S_{0}(p,p'+q')$ and
$S(p-q') \Gamma_{\mu}^{\rm ir}(p-q',p') = S_{0}(p-q') \Gamma_{\mu}
(p-q',p')$, we can identify in
eq.(\ref{jtbornoff}) the piece that contain only $reducible$ vertices and
$free$ Feynman propagators, i.e. the part which exactly coincides with
$J_{\mu \nu}^{B-off}$ from (\ref{jbornoff0}), and a correction term
comming from the mass-operator (see Appendix D):
\be {\tilde{J}_{\mu \nu}^{B-off} \ \ = \ \ J_{\mu \nu}^{B-off} \ \
+ \ \ \delta J_{\mu \nu}^{B-off}},\label{jtbornoff2} \ee
The same procedure for the contact current $\tilde{J}_{\mu \nu}^{C}$
(\ref{jtc}) (see Appendix D) allows us to identify the piece which
contains only $reducible$ vertices and, of course, exactly coincides
with our old contact current $J_{\mu \nu}^{C}$ from (\ref{jc0}) plus
correction term $\delta J_{\mu \nu}^{C}$ caused also by the
mass-operator:
\be {\tilde{J}_{\mu \nu}^{C} \ \ = \ \ J_{\mu \nu}^{C} \ \ + \ \ \delta
J_{\mu \nu}^{C}},\label{jtc2} \ee
The direct calculation of $\delta J_{\mu \nu}^{C}$ gives (see Appendix D):
\be {\delta J_{\mu \nu}^{C} \ \ = \ \ - \delta J_{\mu \nu}^{B-off}
}.\label{djc} \ee
Thus, substituting eqs.(\ref{jtbornoff2},\ref{jtc2}) into eq.(\ref{jtn})
and taking into account eq.(\ref{djc}) we conclude
\be {\tilde{J}_{\mu \nu}^{N} \ \ \ \equiv \ \ \ J_{\mu \nu}^{N}}.
\label{jtn2} \ee
Therefore, in the framework of the ``minimal" scheme for the gauge field
the corrections, connected with the mass-operator, to the off-shell Born
current and to the contact current cancel one another
in the total conserved current. As a result, its definition
in terms of the ir-reducible vertices and full renormalized
propagator as well as in terms of the reducible vertices and
free Feynman propagators is $equivalent$.
\par
Finally, eq.(\ref{jtn2}) is also satisfied if both photons are real (this
can be shown by using the above described procedure with additional
symmetrization$^{14}$ in eq.(\ref{jtc}) on photon lines). Note, the
identity (\ref{jtn2}) leads to some restrictions for the off-shell
effects at the ``photon-point", as a consequence of the gauge invariance.
Indeed, in accordance with eq.(\ref{jgi0}), the right hand side
of eq.(\ref{jtn2}), i.e. current $J^{N}_{\mu \nu}$, does not contain any
off-shellness for the real Compton scattering on a Dirac (or scalar)
particle. This means that the off-shell effects in the left hand side of
eq.(\ref{jtn2}), i.e. in the current $\tilde{J}^{N}_{\mu \nu}$, caused by
the self-energy part, appear in the correction terms only, and cancel in
the total gauge invariant amplitude in a model independent way.
\section{\bf Illustration: Time-Like VCS in ``Unphysical" Region}
As an illustration we consider the possibility to introduce an
on-shell extrapolated Dirac FF in ``unphysical" region and to
investigate it experimentally in the di-lepton photoproduction off
a proton. All information about time-like FF is concentrated in the
VCS amplitude which is much smaller than Bethe-Heitler (BH) one in
general. Therefore, their separation by measuring the $C$- odd $e^{+} e^{-}$-
asymmetry in the $\gamma p \rightarrow p e^{+} e^{-}$ reaction has been
proposed$^{14}$. The selection of ``longitudinal"
photons ${\bf q}^2 \ll q^2 \ll 4M^2$ in the collinear kinematics
provides$^{14}$ a strong suppression of the Pauli FF $F_2$ and the
contribution from resonances, which are excited mainly by magnetic
(transverse) transitions. As a result, for these ``longitudinal" photons
the $physical$ proton resembles
a Dirac one$^{14}$.  Taking into account that squares of BH/VCS
amplitudes are $C$- even under permutation of $e^{+} e^{-}$
($T^{BH,VCS}_{-+} = T^{BH,VCS}_{+-}$), while their interference is $C$-
odd ($T^{int}_{-+} = - T^{int}_{+-}$), one can introduce the $e^{+} e^{-}$-
asymmetry$^{14}$ ($K^{-1}_{-+} = E_{-} + E_{+} \cos(\theta) + p_0$):
$${A_{e^{+}e^{-}} \ \ = \ \ {K^{-1}_{+-} \sigma_{+-} - K^{-1}_{-+}
\sigma_{-+} \over K^{-1}_{+-} \sigma_{+-} + K^{-1}_{-+} \sigma_{-+}} \ \
\equiv \ \ {T^{int}_{-+} \over T^{BH}_{-+} + T^{VCS}_{-+}}
\ \ \approx \ \ T^{int}_{-+} \ / \ T^{BH}_{-+}},$$
where $E_{\pm}$ and $\theta$ are the energy of positron/electron and the
angle between them.  We use c.m. system with z- axis fixed along
3-momentum of the virtual photon ${\bf q}$,
and $\sigma_{+-} = d^5 \sigma_{+-} / d E_{-} d \Omega_{-} d \Omega_{+}$.
\par
The $A_{e^{+}e^{-}}$- asymmetry is shown in fig.1 as a function of
$q^2$ at $E_{\gamma} = 0.65$ GeV, $\theta = 165^o$, $\theta_{\gamma
\gamma^{\star}} = 0^o$ (collinear kinematics with almost ``longitudinal"
photons: ${\bf q}^2 \ll q^2$). To get an idea about the sensitivity of
asymmetry to the $q^2$- dependence, we compare several models for
$F_1(q^2)$ assuming the vector meson dominance (VMD).
The chiral quark-bag model$^{15}$, containing the coupling of the photon to
the meson cloud (VMD piece) and directly to the quark core,
dispersion relation calculations$^{16}$, including constraints from
the hadron-channel unitarity (part corresponding to VMD) and perturbative
QCD, as well as the well known ``dipole fit" were extrapolated
to the time-like ``unphysical" region. All of them practically coincide
and fit experimental data well in the space-like region. When extended
into the time-like sector, they give rise to a resonance-like peak
corresponding to the VMD contribution$^4$, and different results even at
small positive $q^2$. We have also calculated the asymmetry for the
``point-like" proton, when $F_1(q^2) = 1$.
\par
It is interesting to observe that due to the $\pi \pi$- continuum
contribution to
the isovector spectral function$^{16}$, dispersion relation calculations
(VMD + PQCD) predict practically the same asymmetry as a pure meson cloud
piece of the chiral quark-bag model, while the full quark-bag model gives
the result which is very close to the dipole fit prediction. The comparison
of these models with the ``point-like" proton prediction shows a
strong sensitivity of the asymmetry to the EM structure of the proton.
All curves in fig.1 (except the dotted curve) were obtained without
off-shell effects: $F^{+}_1(q^2, W^2) \rightarrow F^{+}_1(q^2, M^2)$, and
expansion (\ref{apl}) was not used at all.
\par
To estimate the size
of off-shell effects (which come only from $J^{GI}_{\mu \nu}$ in a new
representation (\ref{jgi}) of the full conserved current) we
took into account the leading order in the expansion (\ref{ffpl}) on the hadron
virtuality: $a_{11} \neq 0$ and $a_{ik} = 0$, if $i,k > 1$. A small
difference between two lowest curves in the picture, dotted and
double-dashed ones, obtained for the ``point-like" proton, but with $a_{11}
= 1$ and 0, respectively, reflects the insignificant role of off-shell effects
in the full conserved current (\ref{jn0}).  Therefore, selection of ``longitudinal"
photons$^{14}$ in collinear kinematics of $\gamma p \rightarrow p e^{+}
e^{-}$ reaction can provide a practically model-independent
investigation of the ``on-shell extrapolated" Dirac FF in
the ``unphysical" region.
\section{\bf Discussion and Conclusions}
In this paper ``off-shell Born currents" (irregular pieces of the
amplitude) and ``contact currents" (regular pieces of the amplitude) were
obtained ``simul;taniously" from the same principle in an explicit
form, using only gauge properties of the EM field, in terms of the
well-known structure$^{1-4}$ of the half-off-shell vertex $\gamma^{\star}
NN$. So, a completely consistent treatment of irregular
and regular pieces of the total VCS amplitude off a Dirac proton was
achieved without assuming a dynamical model for $off$-$shellness$. This
allows us to consider any re-distributions and cancellations of the
off-shell effects inside the total gauge invariant current in a model
independent way for the first time. Two different representations of the
``minimal" conserved current in terms of the $reducible$ vertices and
$free$ Feynman propagators, as well as $ir$-$reducible$ vertices and
$full$ renormalized propagators were considered and their equivalence was
shown.
\par
To summarize, the definition of the off-shell EM FF through reducible
or irreducible $\gamma^{\star} NN$ vertices is not gauge invariant even
if these vertices satisfy WTI for the 3-point EM Green function.
For the pure EM processes, like VCS,
off-shell effects in the $reducible$ $\gamma^{\star} NN$ vertices may
not only be moved to the regular part of the amplitude (analogously to
the result$^{8.9}$ for $ir$-$reducible$ vertices), but, and it is more
important, they are {\bf cancelled} in the longitudinal
components of the total conserved current by the ``minimal" contact
current, when pole and contact amplitudes are defined consistently.
Therefore, ``charge operators"  are completely free from the off-shellnes
due to gauge constraint. This means that indeed such a model independent
representation of the total conserved current, when off-shell effects
enter only in the transverse (magnetic) gauge invariant $non$-$pole$ part,
exists for both space/time-like regions. This enables one to
introduce an ``on-shell extrapolated" FF for the ``unphysical" region
in a gauge invariant way which could be investigated in practice.
The $unphysical$ FF $F_{1,2}^{-}$ is removed from the full amplitude due to
current conservation and does not influence the observables.
\par
The off-shellness in the ``irreducible" vertices, which
appears even at photon point and which is caused by the mass-operator
corrections, is cancelled in the ``minimal" conserved
current in a model independent way, since self-energy
corrections to the Born and contact currents are the same, but they have
opposite signs. In general this means that the requirements of gauge
invariance impose model independent restrictions on the off-shell
properties of the charge operators at the ``photon point". As a result,
any off-shell effects cannot exist in the real Compton scattering amplitude
off a scalar (or Dirac) particle as a consequence of a gauge invariance
only. Indeed, in this case any off-shellness which appear in the
``irreducible representation" may be canncelled simply by the transition
to the theory in terms of the $reducible$ vertices and $free$ Feynman
propagators.
\par
Let us stress that all above mentioned results were obtained in the
framework of the ``minimal substitution" scheme, on the basis of only
fundamental requirements, such as covariance, CPT and gauge invariance,
without any model suppositions about off-shell behaviour of the nucleon
form factors.
Of course, such a model independent consideration is possible only for
a Dirac or scalar particles when a consistent treatment of the regular
and irregular parts of the total amplitude may be done using only gauge
nature of the EM field. Indeed, ``pure transverse" pieces of the Born and
contact currents (associated with $real$ $photon$), including ``abnormal"
(Pauli) part of the $\gamma NN$ vertex in the one-body current, cannot be
obtained in the ``minimal" scheme only
on the basis of the gauge constraint for 3- and 4-point EM Green
functions (with gauge field). Additional gauge constraint for 4-
and 5-point EM Green functions should be considered on
the level of at least one-loop diagrams$^{10}$ corresponding to
``vertex corrections" to the ``minimal" truncated $\gamma NN$ Green
functions. This evidently requires some knowledge of the strong
interaction structure, i.e. strong dynamics in the
pure hadronic 3- and 4-point Green functions (without gauge field).
However, the off-shellness arising from the ``magnetic" couplings
in the $\gamma^{\star} NN$ vertex (associated with $virtual$ $photon$,
see eq.(\ref{pmredvertex})) may also be moved to the contact current in
a manner similar to eqs. (\ref{ffpl}), (\ref{jbornoff})-(\ref{jnongi}).
\par
Finally, measurements of $A_{e^{+}e^{-}}$- asymmetry
in the $\gamma p \rightarrow p e^{+} e^{-}$ reaction with ``longitudinal"
photons, with a mass much more larger than 3-momentum, can be used to
investigate the on-shell extrapolated Dirac
FF $F_1^{+}(q^2,M^2)$ of the proton in the time-like region
$4m^2_e <q^2 < 4M^2$.
\par
When this work was completed, some attempt to estimate the role of
off-shell effects in the VCS was published$^{17}$.
\acknowledgments
This work was supported by the Nederlandse Organisatie voor
Wetenschappelijk Onderzoek (NWO) and Fundamenteel Onderzoek der
Materie (FOM).
\section{\bf Appendix A}
The general form of the Ward-Takahashi identity for the $reducible$
three-point Green
function with the EM field (vertex $\gamma NN$ in our case) is well known
$${q_{\mu} \Gamma_{\mu}(p,p') = e S_{0}^{-1}(p)[S(p') - S(p)] S_{0}^{-1}(p')}.
\eqno(A.1)$$
Using the asymptotic relation
$\lim_{p^2 \rightarrow M^2} S_{0}^{-1}(p) S(p) = 1 $
the gauge constraint for the $half$-$off$-$shell$ reducible vertex becomes
(let us put $p$ is an on-shell momentum, while $p'$ is an
off-shell one: $p^2=M^2$, $p'^2 \neq M^2$, and $u(p)$ is a Dirac
bi-spinor, i.e. a wave function of the on-shell particle):
$${\bar{u}(p) q_{\mu} \Gamma_{\mu}(p,p') = e \bar{u}(p) S_{0}^{-1}(p)
[S(p') - S(p)] S_{0}^{-1}(p') = e \bar{u}(p) \{S_{0}^{-1}(p) -
S_{0}^{-1}(p') \}}\eqno(A.2)$$
Therefore from eqs.(A.1) and (A.2) we see that the simple form of
eq.(\ref{wtib}) for the Ward-Takahashi identity may be used for the
half-off-shell $\gamma NN$ vertex.
\section{\bf Appendix B}
The structure of the VCS $off$-$shell$ $Born$ current in
eq.(\ref{jbornoff0}) will be
considered here on  the basis of the ``minimal insertion" of the EM
field.  We start with the definition of the reducible $\gamma^{\star} NN$
vertex $\Gamma_ {\mu}(p',p)$ in terms of the  irreducible
$\Gamma^{\rm ir}_{\mu}(p',p)$ one, eq.(\ref{eqno(1)}).
\par
Since the $reducible$ vertex in eq.(\ref{eqno(1)}) is surrounded by free propagators,
we have to deal with the ``minimal insertion" of the external EM field
$A_{\nu}(q')$, corresponding to the real (second) photon,
into the ``free" external hadron lines in the right-hand side of
eq.(\ref{eqno(1)})
$${S_{0}(p') \Gamma_{\mu}(p',p) S_{0}(p) \ \ \rightarrow
\ \ S_{0}(p'-e A) \Gamma_{\mu}(p',p) S_{0}(p-e A)
\equiv G_{\mu}(p',p,\{ A \})
}\eqno(B.1)$$
Then, the corresponding EM current is:
$${J_{\mu \nu}^{B-off} = S^{-1}_{0}(p') \ \ [{\delta \over
\delta A_{\nu}}
G_{\mu}(p',p,\{A\}) ]_{A \rightarrow 0} \ \ S_{0}^{-1}(p)},\eqno(B.2)$$
Using (B.1) and (B.2) we obtain the off-shell Born current for VCS in
the form of eq.(\ref{jbornoff0}).
\par
The same result may be obtained if one takes into
account that the ``minimal" three-point Green function, corresponding to
the ``minimal" coupling of the real photon to the external hadron lines,
is connected with the corresponding inverse propagator$^{10}$ (in our
case, see eq.(\ref{eqno(1)}), it is the $free$ propagator):
$$\Gamma_{\nu}(p-q',p) = - e \int^{1}_{0} {d \lambda \over \lambda}
{d \over d q_{\nu}} S_{0}^{-1}(p-\lambda q') = e \gamma_{\nu},$$
\section{\bf Appendix C}
A general Lorentz covariant, crossing symmetry, CPT and gauge
invariant expression of
the $off$-$shell$ nucleon current for the space/time-like VCS,
containing independently crossing symmetry off-shell Born term
and contact current both in an explicite form as a function of only
half-off-shell nucleon form factors in the $\gamma^{\star} NN$ vertex,
was obtained recently$^{14}$.
\par
In the particular case, when one photon is
$real$ and the other  $virtual$, the crossing transformation connects the
amplitudes of the $time$-$like$ and the $space$-$like$ VCS. Indeed from
eqs.(\ref{jbornoff0},\ref{jc0}) a total gauge invariant current for
time-like VCS ($q^2>0$, $q'^2=0$) is:
$${e^{-1} J^{N}_{\mu \nu}(\gamma N \rightarrow N \gamma^{\star}) =
\Gamma_{\mu}(p,p'+q') S_{0}(p'+q') \gamma_{\nu} + \gamma_{\nu} S_{0}(p-q')
\Gamma_{\mu}(p-q',p') }$$
$${- \int^1_0 {d \lambda \over \lambda}
{d \over d q'_{\nu}} \{ \Gamma_{\mu}(p, p'+\lambda q') - \Gamma_{\mu}
(p-\lambda q',p') \} }.\eqno(C.1)$$
Equation (C.1) under crossing transformation $q \rightarrow -q'$, $q'
\rightarrow -q$, $\mu \leftrightarrow \nu$ generates
a gauge invariant current for space-like VCS ($q^2=0$, $q'^2<0$):
$${e^{-1} J^{N}_{\nu \mu}(\gamma^{\star} N \rightarrow N \gamma) =
\Gamma_{\nu}(p,p'-q) S_{0}(p'-q) \gamma_{\mu} + \gamma_{\mu} S_{0}(p+q)
\Gamma_{\nu}(p+q,p') }$$
$${- \int^1_0 {d \lambda \over \lambda}
{d \over d q'_{\mu}} \{ \Gamma_{\nu}(p+\lambda q,p') - \Gamma_{\nu}
(p,p'-\lambda q) \} },\eqno(C.2)$$
Although the currents (C.1) and (C.2) are not the same (since the initial
and final photons are different), they are connected by crossing
transformations.
\par
At the photon point (both photons are real) due to WTI(\ref{wtib})
all reducible vertices $\Gamma_{\mu}(p,p')$ in eqs.(\ref{jbornoff},\ref{jc0})
become simply a constant ($ e \gamma_{\mu}$). This
means that for real photons the reducible half-off-shell $\gamma pp$ vertex
of the Dirac proton coincides with the on-shell one and does not  contain
any $off$-$shellness$ (consequence of the gauge constraint$^{2,10}$).
As a result, $J_{\mu \nu}^{B-off}(q^2=q'^2=0) =
J_{\mu \nu}^{B-on}(q^2=q'^2=0) \ \ , J_{\mu \nu}^{C}(q^2=q'^2=0) = 0$,
and in the case of two real photons current
(\ref{jn}) is crossing symmetry
$$ J^{N}_{\mu \nu}(P, q', q)|_{q^2=q'^2=0} =
J^{N}_{\nu \mu}(P, -q, -q')|_{q^2=q'^2=0} $$
\section{\bf Appendix D}
First, we derive the connection between
two different off-shell Born currents $\tilde{J}_{\mu \nu}^{B-off}$
and $J_{\mu \nu}^{B-off}$, defined in terms of the $ir$-$reducible$
vertices and $full$ renormalized propagators, and in terms of the
$reducible$ vertices and $free$ Feynman propagators, respectively.
Starting from eq.(\ref{jtbornoff}), and accounting for the definition
(\ref{eqno(1)}), we can make transition from the irreducible
vertices of the virtual photons and full propagators to the reducible
ones and free propagators, while for the real photons we have still
irreducible vertices:
$${\tilde{J}_{\mu \nu}^{B-off} = \Gamma_{\mu}(p,p'+q') S_{0}(p'+q')
\Gamma_{\nu}^{ir}(p'+q'.p') + \Gamma_{\nu}^{ir}(p,p-q') S_{0}(p-q')
\Gamma_{\mu}(p'+q'.p')}.$$
The substitution of eq.(\ref{gammair3}) for irreducible $\gamma NN$ vertices
allows us to decompose the ``new" off-shell Born current (\ref{jtbornoff})
into two terms. The first one is the ``old"
Born current $J_{\mu \nu}^{B-off}$ from eq.(\ref{jbornoff0}), while
the second one, $\delta J_{\mu \nu}^{B-off},$ is a correction term coming
from the mass-operator:
$${\tilde{J}_{\mu \nu}^{B-off} = \Gamma_{\mu}(p,p'+q') S_{0}(p'+q')
(e \gamma_{\nu})+(e \gamma_{\nu}) S_{0}(p-q') \Gamma_{\mu}(p-q'.p')
+ \delta J_{\mu \nu}^{B-off} },\eqno(D.1)$$
$${\delta J_{\mu \nu}^{B-off} = - e \int_{0}^{1} {d \lambda \over
\lambda} \{ \Gamma_{\mu}(p,p'+q') S_{0}(p'+q')
[{d \over d q'_{\nu}}
\Sigma(p'+\lambda q')] - }$$
$${-
[{d \over d q'_{\nu}}
\Sigma (p-\lambda q')]
S_{0}(p-q') \Gamma_{\mu}(p-q',p') \} }.
\eqno(D.2)$$
From eqs.(D1,2), and using eq.(\ref{gammair3}), we immediately get
the relation between two off-shell Born currents in the form of
eq.(\ref{jtbornoff2}).
\par
Second, we  derive the connection between two the different contact
currents $J_{\mu \nu}^{C}$ and $\tilde{J}_{\mu \nu}^{C}$ which are
defined through irreducible and reducible vertices, respectively.
Starting from eq.(\ref{jtc}), we can rewrite it as
$${\tilde{J}^{C}_{\mu \nu} = - e \int^1_0 {d \lambda \over \lambda}
{d \over d q'_{\nu}} \{ \Gamma_{\mu}^{\rm ir}(p, p'+\lambda q') S(p'+
\lambda q') S^{-1}(p'+\lambda q') - }$$
$${- S^{-1}(p-\lambda q') S(p-\lambda q') \Gamma_{\mu}^{\rm ir}
(p-\lambda q',p') \} }.\eqno(D.3)$$
Using the definition (\ref{eqno(1)}), and the relation between $free$
and $full$ propagators (\ref{smin1}), we can make the transition
from the irreducible vertices to the reducible ones:
$${\tilde{J}^{C}_{\mu \nu} = - e \int^1_0 {d \lambda \over \lambda}
{d \over d q'_{\nu}} \{ \Gamma_{\mu}(p, p'+\lambda q') S_{0}(p'+
\lambda q') [S_{0}^{-1}(p'+\lambda q') - \Sigma(p'+\lambda q')] - }$$
$${- [S_{0}^{-1}(p-\lambda q') - \Sigma(p-\lambda q') S_{0}(p-\lambda q')
\Gamma_{\mu}(p-\lambda q',p') \} }.\eqno(D.4)$$
Isolating in (D.4) terms containing only reducible vertices, we see
that the ``new" contact current (\ref{jtc}) may be presented in the form
(\ref{jtc2}), i.e.
expressed through the ``old" contact current $J_{\mu \nu}^{C}$, defined by
eq.(\ref{jc0}), and an additional correction term $\delta J_{\mu \nu}^{C}$:
$$\delta J_{\mu \nu}^{C} =  e \int^1_0 {d \lambda \over \lambda} \{
[{d \over d q'_{\nu}}
\Gamma_{\mu}(p, p'+\lambda q')] S_{0}(p')
\Sigma(p') + \Gamma_{\mu}(p, p'+ q') [{d \over d q'_{\nu}} S_{0}(p'+
\lambda q')] \Sigma(p') + $$
$$+ \Gamma_{\mu}(p, p'+ q') S_{0}(p'+ q') [{d \over d q'_{\nu}}
\Sigma(p'+\lambda q')] - [{d \over d q'_{\nu}} \Sigma(p-\lambda q')]
S_{0}(p-q') \Gamma_{\mu} (p-q',p') - $$
$${- \Sigma(p) [{d \over d q'_{\nu}}
S_{0}(p-\lambda q')]  \Gamma_{\mu} (p-q',p') - \Sigma(p) S_{0}(p)
[{d \over d q'_{\nu}} \Gamma_{\mu} (p-q',p')] \}}.\eqno(D.5)$$
Then, taking into account (\ref{mass}), and omitting all terms which are
proportional to $\Sigma(p)$ and $\Sigma(p')$ since
$p^2=p'^2=M^2$, we can write down the ``contact current correction":
$${\delta J_{\mu \nu}^{C} =  e \int^1_0 {d \lambda \over \lambda}
\{\Gamma_{\mu}(p, p'+ q') S_{0}(p'+ q')
[{d \over d q'_{\nu}}
\Sigma(p'+\lambda q')] - }$$
$${
[{d \over d q'_{\nu}}
- \Sigma(p-\lambda q')]
S_{0}(p-q')
\Gamma_{\mu} (p-q',p') \} }.\eqno(D.6)$$
Comparing eqs. (D.2) and (D.6), we immediately get eq.(\ref{djc}).

\bigskip
{\bf Figure Captions}
\\  \\
fig.1  \\  Predictions for the asymmetry $A_{e^{+}e^{-}}$
for different models for $F_1(q^2,M^2)$.

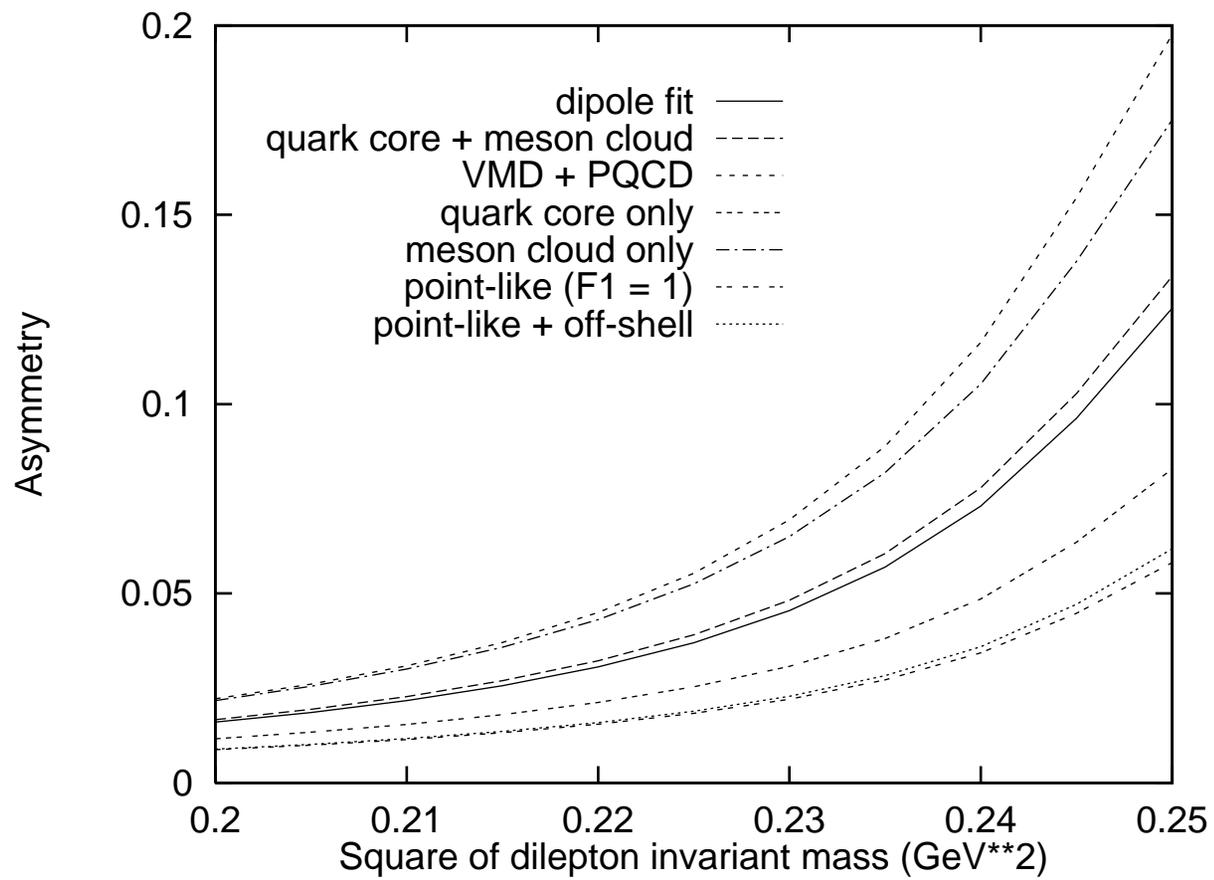
\end{document}